\documentclass[aps,prl,reprint,10pt,floatfix,showpacs,superscriptaddress,showpacs]{revtex4-1}
\usepackage{graphicx}
\usepackage{amsmath}
\usepackage{amsfonts}
\usepackage{hyperref}
\usepackage{enumitem}
\usepackage{color}

\footnotetext[3]{The two authors contributed  equally to this work.}
\newif\ifWithAppendix
\WithAppendixtrue

\begin{document}

\title{Comment on ``Solving Statistical Mechanics Using VANs'': \\ Introducing saVANt---VANs Enhanced by Importance and MCMC Sampling}

\author{Kim Nicoli\textsuperscript{$\ddagger$}}
\affiliation{Machine Learning Group, Technische Universit\"{a}t Berlin, 10587 Berlin, Germany}

\author{Pan Kessel\textsuperscript{$\ddagger$}}
\affiliation{Machine Learning Group, Technische Universit\"{a}t Berlin, 10587 Berlin, Germany}
\affiliation{Berlin Big Data Center, 10587 Berlin, Germany}

\author{Nils Strodthoff}
\affiliation{Fraunhofer Heinrich Hertz Institute, 10587 Berlin, Germany}

\author{Wojciech Samek}
\email{wojciech.samek@hhi.fraunhofer.de}
\affiliation{Fraunhofer Heinrich Hertz Institute, 10587 Berlin, Germany}
\affiliation{Berlin Big Data Center, 10587 Berlin, Germany}
\affiliation{Berliner Zentrum f\"ur Maschinelles Lernen, 10587 Berlin, Germany}

\author{Klaus-Robert M\"{u}ller}
\email{klaus-robert.mueller@tu-berlin.de}
\affiliation{Machine Learning Group, Technische Universit\"{a}t Berlin, 10587 Berlin, Germany}
\affiliation{Berlin Big Data Center, 10587 Berlin, Germany}
\affiliation{Department of Brain and Cognitive Engineering, Korea University, Anam-dong, Seongbuk-gu, Seoul 136-713, South Korea}
\affiliation{Max-Planck-Institut f\"{u}r Informatik, Saarbr\"{u}cken, Germany}
\affiliation{Berliner Zentrum f\"ur Maschinelles Lernen, 10587 Berlin, Germany}

\author{Shinichi Nakajima}
\affiliation{Machine Learning Group, Technische Universit\"{a}t Berlin, 10587 Berlin, Germany}
\affiliation{Berlin Big Data Center, 10587 Berlin, Germany}
\affiliation{RIKEN Center for AIP, 103-0027, Tokyo, Japan}



\maketitle

In statistical physics and strongly-interacting field theories, physical observables are often computed using Monte-Carlo methods; most interestingly near criticality. 
In a recent exciting publication \cite{PhysRevLett.122.080602}, 
the authors have demonstrated the promising potential of generative neural samplers (GNSs)
in this context. 
In this comment, we propose a subtle yet crucial modification which enhances their approach drastically by giving theoretical guarantees and fast computation. We envision our method to be helpful for broad applications in high-energy and condensed matter physics.

The \emph{Variational Autoregressive Network} (VAN)  \cite{PhysRevLett.122.080602}
is a GNS, successfully trained so as to provide \emph{independent} samples from an approximate distribution $q(s)$ of the Boltzmann distribution $ p(s)= 1/Z \, \exp(-\beta H(s))$.
As the authors showed, thermodynamic observables (expectation values of an operator $\mathcal{O}$) can be estimated by the sample mean $\langle \mathcal{O}(s) \rangle_{p} \approx \frac{1}{N}\sum_{i} \mathcal{O}(s_i)$ of the samples  drawn from a VAN $s_i \sim q $.

Our proposal, the \emph{sampled VAN training} (saVANt) is motivated by an exceptional feature of VANs, namely that the exact sampling probability $q(s)$ is accessible unlike in most of the popular GNSs \cite{Goodfellow14,Kingma14,energy,rbm}. 
Taking advantage of this fact, we posit that the sampling error can be corrected by using neural network-based MCMC or importance sampling which leads to \emph{asymptotically unbiased} estimators for physical quantities.

Practically, we train saVANt samplers using the same approach as Wu et al.\,but with output probabilities bounded within $[\epsilon, 1 - \epsilon]$ for small $\epsilon >0$. Then, \emph{Neural Importance Sampling} (saVANt-NIS) defined below gives an unbiased estimator:
\begin{align}
    \langle \mathcal{O}(s) \rangle_{p} \approx 
    \textstyle \sum_{i} w_i \, \mathcal{O}(s_i) &&\text{with} &&  s_i \sim q  \,, \nonumber
\end{align}
where $w_i = \tfrac{\hat{w}_i}{\sum_i \hat{w}_i}$ for $\hat{w}_i = \tfrac{e^{-\beta H(s_i)}}{q(s_i)}$
is the importance weight.
Alternatively, \emph{Neural MCMC Sampling} (saVANt-NMCMC) 
with the acceptance probability
\begin{align}
\text{min}\left(1, \tfrac{p_0(s'| s) \, p(s')}{p_0(s| s') \, p(s)}\right)
    = \text{min}\left(1, \tfrac{q(s') \, \exp(-\beta H(s'))}{q(s) \, \exp(-\beta H(s))}\right)
    \nonumber
\end{align}
also provides an unbiased estimator.
Note that, in saVANt-NMCMC, we can draw samples from the sampler \emph{independently}, and apply the Metropolis rejection step to the samples arranged in an \emph{arbitrary} order.
This is
because the trial distribution $p_0(s'| s) = q(s')$ is independent from the previous sample $s$.  

The asymptotic unbiasedness is guaranteed even if $q$ does not accurately approximate the target distribution $p$ (see e.g.\ \cite{murphy}),
which allows \emph{transfer across parameter space}. For example,
a sampler trained at a certain temperature can be used to estimate physical observables at other temperatures.
\begin{figure}[t] 
\includegraphics[width=0.5\textwidth]{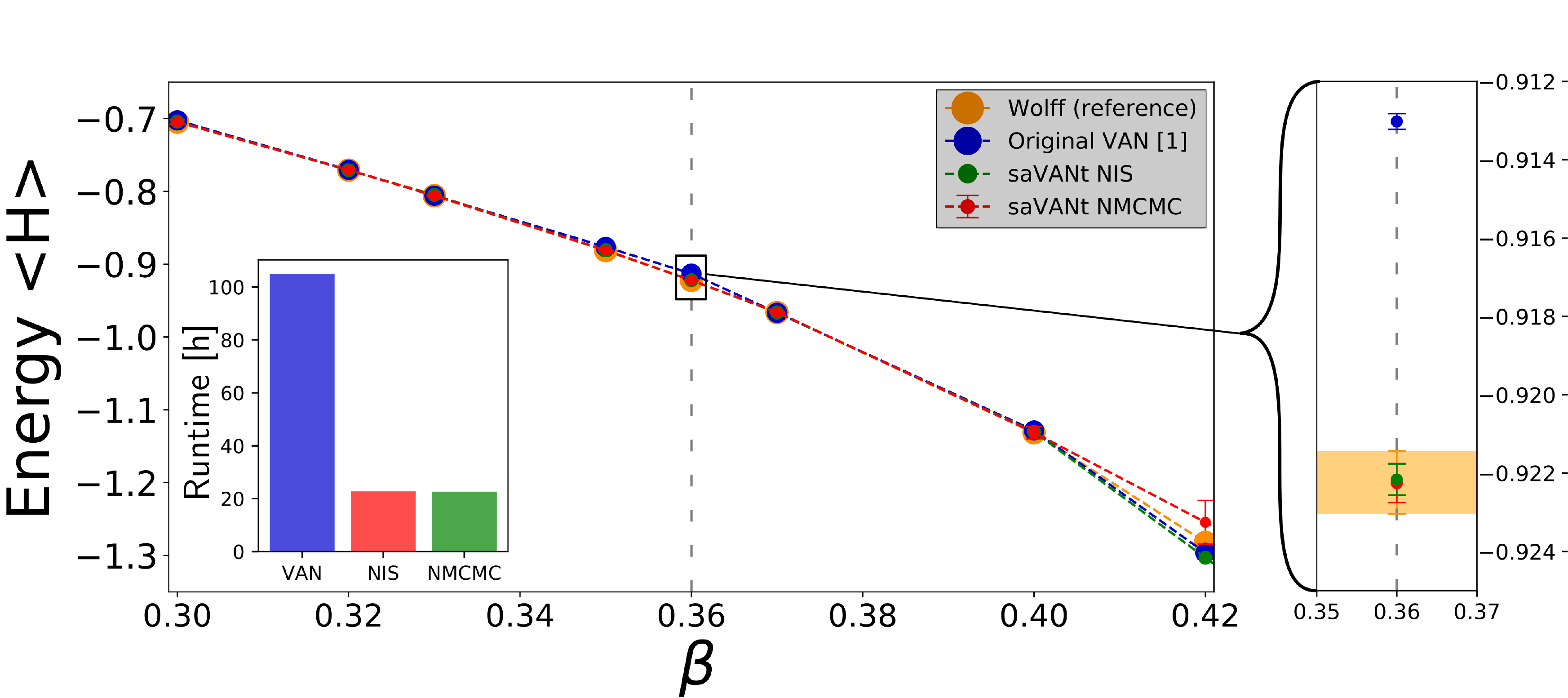}
\vspace{-5mm}
\caption{\label{fig:main} Original VAN requires to train a sampler for every $\beta$ (8 VANs were trained independently), while saVANt trains a single sampler at $\beta=0.36$ and uses it for probing 8 different $\beta$. This reduced the runtime significantly (see the bar graph at the left-hand side). Furthermore, our unbiased estimator is significantly more accurate than the original VAN and is compatible with the Wolff algorithm (see the right most graph).}
\vspace{-3mm}
\end{figure}
We demonstrate the effectiveness of our proposal for the Ising model with $H(s)=- \sum_{<i, j>} s_i s_j$. We trained all samplers on a Tesla P100 GPU by using the implementation provided by Wu et al \cite{PhysRevLett.122.080602}. The error bars were determined using the method described in \cite{uwerr}, and all runs used $500$k samples. 
Figure~\ref{fig:main} clearly shows that saVANt with a single trained sampler accurately predicts the energies for \emph{different} temperatures and reduces the runtime considerably.  We also observe that the estimator by original VAN is not fully compatible with the reference value provided by the Wolff algorithm \cite{uwalg}, whereas saVANt is.

For analyzing complex physical systems,
the unbiasedness of the estimators is of particular importance 
since any GNS can hardly be expected to capture the underlying physics perfectly at reasonable training costs.  Practically, the saVANt framework could allow to predict observables close to the critical point based on samplers trained further away from it. More importantly, we consider saVANt a viable strategy to explore expensive parts in parameter space based on models trained in cheaper regions. Furthermore, our saVANt-NMCMC can be easily combined with well-established Monte-Carlo methods as a generalized overrelaxation step. We believe that this may alleviate the local minima problem and large autocorrelation times---key issues in
studying critical phenomena and the continuum limit of lattice field theories.

\textit{Acknowledgment--}
We thank Wu et al.\ for providing their implementation. This work was supported by the German Ministry for Education and Research as Berlin Big Data Center (01IS18025A) and Berlin Center for Machine Learning (01IS18037I). This work was also supported by the Information \& Communications Technology Planning \& Evaluation (IITP) grant funded by the Korea government (No. 2017-0-00451).

\bibliography{comment}

\ifWithAppendix

\onecolumngrid

\newpage

\appendix

\section{Additional Details on Algorithm}
\subsection{Lightning Review of VAN}
Wu~et~al approximate Boltzmann distributions $p(s)=\frac{1}{Z} \, \exp(-\beta H(s) )$ with an auto-regressive generative model $q$ by minimizing the KL divergence
\begin{align}
    \text{KL}( q | p) &= \sum_s q(s) \ln{\frac{q(s)}{p(s)}} = \sum_s q(s) \left(  \ln{q(s)} + \beta H(s) + \beta Z \right) \,.
\end{align}
A PixelCNN is used which allows exact evaluation of the probability $q(s)$ and relatively efficient sampling. Also note that the partition function $Z$ is a constant and  therefore the last summand leads to no contribution to the gradient. As a result, VAN can be trained by sampling from the model $q$, evaluating the probability $q(s)$ and the Hamiltonian $H(s)$ for these samples and using gradient descent.  

\subsection{Bounding the output probabilities of VAN}

We can simply interpret the original network output $q' \in [0, 1]$ as the probability $q \in [\epsilon, 1-\epsilon]$ by the following mapping:
\[
q = \left(q' - \frac{1}{2}\right) \left(1 - 2 \epsilon\right) + \frac{1}{2}.
\]

\subsection{Proof: Estimators are Asymptotically Unbiased}
Assume that the support of the sampling distribution $q$ contains the support of the target distribution $p$. This property is ensured by ensuring that the probability takes values in $q \in [\epsilon, 1-\epsilon]$.
\subsubsection{Neural Importance Sampling}
Then, importance sampling with respect to $q$, i.e. 
\begin{align}
    \langle \mathcal{O}(s) \rangle \approx \sum_{i=1}^N w_i \, \mathcal{O}(s_i) \,, &&  s_i \sim q(s) \,, && w_i = \frac{\hat{w}_i}{\sum_i \hat{w}_i} \,, && \, \hat{w}_i = \frac{e^{-\beta H(s_i)}}{q(s_i)} \,.
\end{align}
is an asymptotically unbiased estimator of the expectation value $\langle \mathcal{O}(s) \rangle$ because
\begin{align}
    \langle \mathcal{O}(s) \rangle_p = \sum_s p(s) \mathcal{O}(s) = \sum_s q(s) \, \frac{p(s)}{q(s)} \,  \mathcal{O}(s) = \frac{1}{Z}  \sum_s q(s)\, \underbrace{\frac{\exp(-\beta H(s))}{q(s)}}_{=\hat{w}(s)} \,  \mathcal{O}(s) \approx \frac{1}{Z N}  \sum_{i=1}^N \hat{w}(s_i) \mathcal{O}(s_i) \,,
\end{align}
where $s_i \sim q$. The partition function $Z$ can be similarly determined
\begin{align}
    Z = \sum_s \exp(-\beta H(s)) = \sum_s q(s) \frac{\exp(-\beta H(s))}{q(s)} \approx \frac{1}{N} \sum_{i=1}^N  \hat{w}(s_i) \,.
\end{align}
Combining the previous equations, we obtain 
\begin{align}
 \langle \mathcal{O}(s) \rangle_{p} \approx \sum_{i} w_i \, \mathcal{O}(s_i) && \text{with} && w_i = \frac{\hat{w}_i}{\sum_i \hat{w}_i} \,.
\end{align}

\subsubsection{Neural MCMC Sampling}
The sampler $q$ can be used as a trial distribution $ p_0(s' | s) = q(s')$ for a Markov-Chain which uses the following acceptance probability in its Metropolis step
\begin{align}
    p_a(s'| s) = \text{min}\left(1, \frac{p_0(s| s') p(s')}{p_0(s'| s) p(s)}\right)  = \text{min}\left(1, \frac{q(s) \, \exp(-\beta H(s'))}{q(s') \, \exp(-\beta H(s))}\right)  \,. 
\end{align}
This fulfills the detailed balance condition
\begin{align}
    p_t(s'|s) \exp(-\beta H(s)) = p_t(s|s') \exp(-\beta H(s))
\end{align}
because the total transition probability is given by $p_t(s'|s)=q(s') \, p_a(s'|s)$ and therefore
\begin{align*}
     p_t(s'|s) \exp(-\beta H(s)) &= q(s') \,  \text{min}\left(1, \frac{q(s) \, \exp(-\beta H(s'))}{q(s') \, \exp(-\beta H(s))}\right) \,  \exp(-\beta H(s)) \\ &= \text{min}\left\{q(s') \exp(-\beta H(s)), \, q(s)\exp(-\beta H(s'))\right\} \\
     &= p_t(s|s') \exp(-\beta H(s')) \,,
\end{align*}
where we have used the fact that the min operator is symmetric and that all factors are strictly positive. The latter property is ensured by the fact that $q(s) \in [\epsilon, 1-\epsilon]$.

\section{Additional Details on Experiments}
\subsection{Setup}
We use a Tesla P100 GPU with 16GB of memory both for training and sampling. A $16\times16$ lattice is considered. The reference values are generated using the Wolff algorithm using 2M steps with 100k warm-up steps. We use the ResNet version of VAN with the following hyperparameter choices (chosen to match with the ones used by Wu et al.):

\begin{table}[h!]
\centering
\begin{tabular}{||c c ||} 
 \hline
name &  value \\ 
\hline
net depth & 6 \\
net width & 3 \\
half kernel size & 3 \\ 
bias & true \\
epsilon & 1e-07 \\
 \hline
\end{tabular}
\caption{Hyperparameters for VAN used in our experiments.}
\label{table:trap}
\end{table}
The reference implementation of Wu et al. is used to train the VANs. For estimating the results of VAN and saVANt-NIS, we use 1000 iterations sampling 500 configurations each. For saVANt-NMCMC, we use 100k steps sampling 500 candidate configurations in a batch. No warm-up steps are required because candidates are sampled from a pre-trained VAN. As is demonstrated in Table~\ref{table:runtime}, all algorithms have roughly the same runtime for sampling but saVANt leads to significant reduction in training time as explained in the main text.

\subsection{Additional Results}
\subsubsection{Properties of saVANt-NMCMC}
saVANt-NMCMC is less likely to get stuck in local minima and has reduced autocorrelation times as compared to Metropolis. This is demonstrated in Table~\ref{table:trap} and \ref{table:tint}. 
\begin{table}[h!]
\centering
\begin{tabular}{||c | c | c||} 
 \hline
Metropolis &  saVANt NMCMC & VAN \\ [0.5ex] 
 \hline
-0.99 $\pm$ 2e-5 & -4e-4 $\pm$ 3e-3 & -6e-4 $\pm$ 2e-3   \\ 
 \hline
\end{tabular}
\caption{Magnetization $M$ at $\beta=1.0$. By $\mathbb{Z}_2$-symmetry, the correct result has to be close to vanishing. In contrast to the Metropolis algorithm, both saVANt and VAN can jump in configuration space between (regions close to) the two degenerate minima of the Ising model.}
\label{table:trap}
\end{table}

\begin{table}[h!]
\centering
\begin{tabular}{||c | c | c||} 
 \hline
$\beta$ & Metropolis &  saVANt-NMCMC \\ [0.5ex] 
 \hline
0.44  &  231.1 $\pm$ 14.0   &  0.50 $\pm$ 0.01   \\ 
0.45  &   542.5 $\pm$ 47.4  &  0.50 $\pm$ 0.01   \\ 
 \hline
\end{tabular}
\caption{Integrated autocorrelation time $\tau_{M, int}$ of the magnetization for $\beta=0.44$ and $\beta=0.45$. Both runs use 5M steps. }
\label{table:tint}
\end{table}
\subsubsection{Using Models Trained at Different Temperatures}

As already shown in the main text, Fig.~\ref{fig:beta57} and Fig.~\ref{fig:beta48} demonstrate that the bias for VANs holds for different betas. In particular, we looked at $\beta=0.48$ and $\beta=0.57$. As for the main study shown in the manuscript for $\beta=0.36$, here we trained a VAN with the aforementioned setup at the reference $\beta$, and we subsequently used this model to predict energies at different $\beta$. We note that VAN does not reproduce the reference values for almost all values of $\beta$. This discrepancy is particularly pronounced as one approaches the critical temperature. Fig. \ref{fig:fixed reference beta} shows the aforementioned transfer property from a different perspective.

\begin{table}[h!]
\centering
\begin{tabular}{||c c c||} 
 \hline
VAN &  saVANt-NMCMC & saVANt-NIS \\ [0.5ex] 
 \hline\hline
 1h 7m & 1h 7m & 1h 8m  \\ 
 \hline
\end{tabular}
\caption{Runtime for 500k configurations on a Tesla P100 GPU with 16GB memory. This does not include the time for training which is significantly lower for saVANt as compared to VAN.}
\label{table:runtime}
\end{table}

\begin{figure}[h!]
\includegraphics[width=1.0\textwidth]{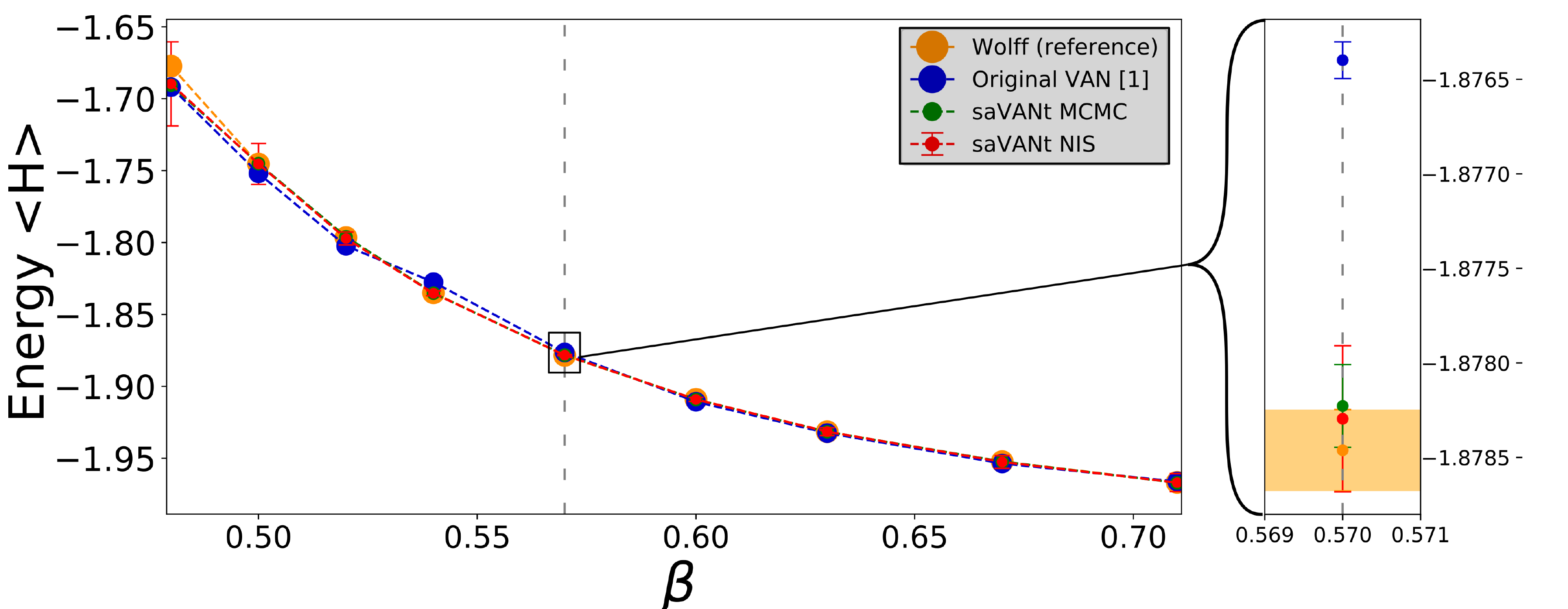}
\caption{\label{fig:beta57} Analogous plot to the main plot in the paper but for $\beta=0.57$.}
\end{figure} 

\begin{figure}[h!]
\includegraphics[width=1.0\textwidth]{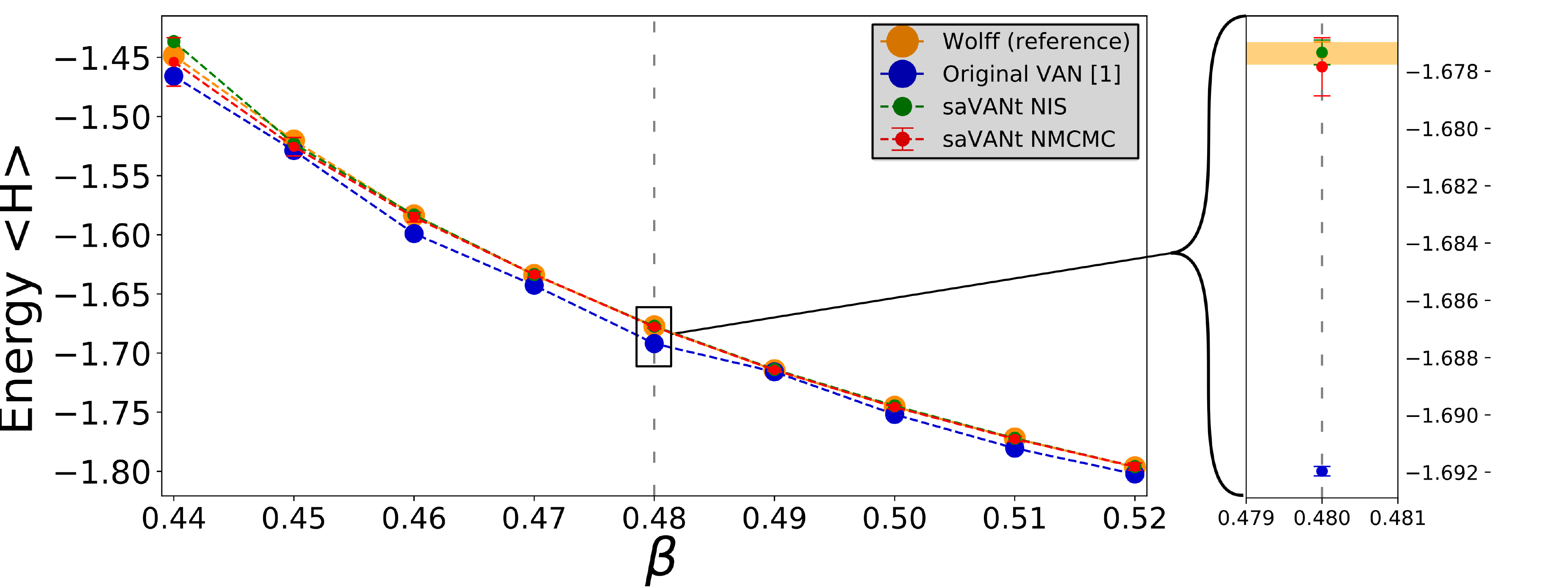}
\caption{\label{fig:beta48} Analogous plot to the main plot in the paper but for $\beta=0.48$ which closer to the inverse critical temperature. Note that the sampling error of VAN gets more pronounced in this regime.}
\end{figure} 

\begin{figure}[h!]
\includegraphics[width=1.0\textwidth]{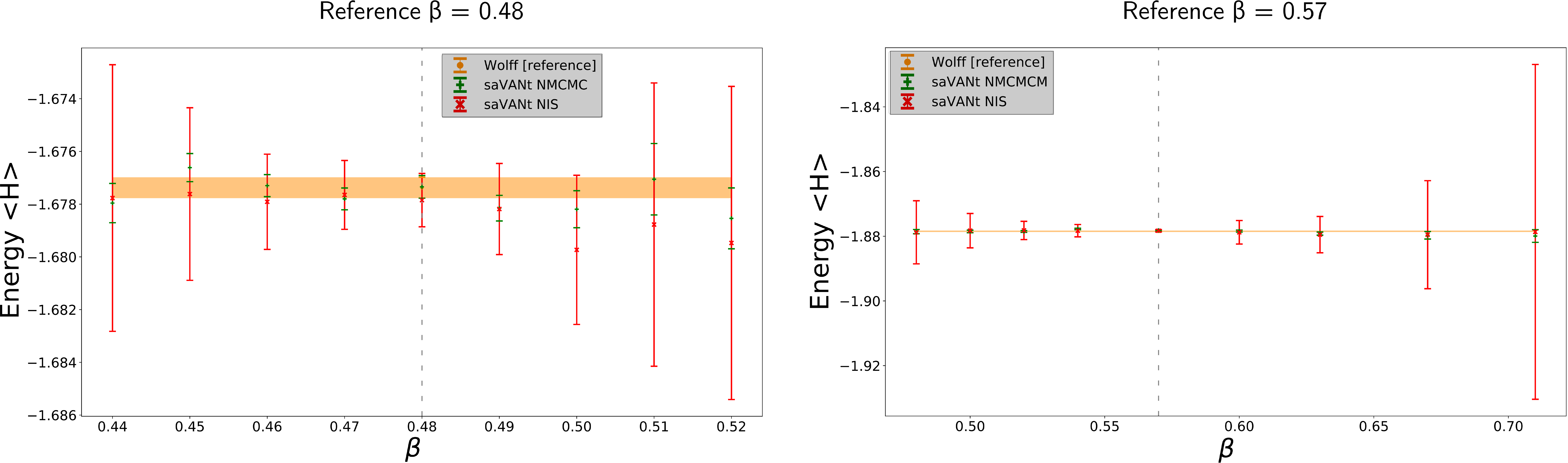}
\caption{\label{fig:fixed reference beta} In this plot, we show the transfer property for two reference value for $\beta$. The horizontal axis denotes the $\beta$ value used for training unlike in the previous plots. We trained samplers at many different betas. Then we estimate energies at a reference value (0.48 on the left and 0.57 on the right) using both saVANt-MCMC and saVANt-NIS. The orange line shows the estimate of the energy at the reference $\beta$ as determined by the Wolff algorithm.}
\end{figure}

\fi
\end{document}